%% file: vdtak_ffs.tex
\documentclass{iaus}
\usepackage{graphicx}
\usepackage{natbib}
\include{defs}
\title[Chemistry of high-mass star formation] 
{The chemistry of high-mass star formation}

\author[Van der Tak]   
{Floris F.\ S.\ van der Tak}
\affiliation{Max-Planck-Institut f\"ur Radioastronomie, Auf dem H\"ugel 69,
  53121 Bonn, Germany \break e-mail: vdtak@mpifr-bonn.mpg.de}

\pubyear{2005}
\volume{227}
\pagerange{}
\date{May 16-20, 2005}
\jname{Massive Star Birth: A Crossroads of Astrophysics}
\editors{R.\ Cesaroni \& C.M.\ Walmsley, eds.}
\begin{document}

\maketitle

\begin{abstract}
  This paper reviews the chemistry of star-forming regions, with an emphasis on
  the formation of high-mass stars. We first outline the basic molecular
  processes in dense clouds, their implementation in chemical models, and
  techniques to measure molecular abundances. Then, recent observational,
  theoretical and laboratory developments are reviewed on the subjects of hot
  molecular cores, cosmic-ray ionization, depletion and deuteration, and oxygen
  chemistry.  The paper concludes with a summary of outstanding problems and
  future opportunities.
\end{abstract}



\section{Introduction: Why astrochemistry?}
\label{s:intro}

The cool and dense material in the interstellar medium of galaxies plays an
essential role in the life cycle of stars: it is the place where new stars
are born. The only way to probe this fundamental process is through observations
of molecular lines and of dust continuum at infrared and \smm\ wavelengths. 
The interpretation of such observations requires understanding of chemical
processes for several reasons. The chemical composition of star-forming matter
is very sensitive to physical parameters such as temperature, density, and
ionization rate. Molecular observations thus serve as probes of physical
processes such as grain growth, shocks, or cosmic-ray impact.  \textit{Vice versa},
observations of star formation may be aimed at a particular physical component
such as the disk or the outflow by choosing lines of particular molecules.
Furthermore, the chemical composition of the clouds is a strong function of
time.  If the temperature and density structure of a region are known, its
chemical age may be estimated by comparing its observed chemical composition to
model calculations.  In addition, the molecular content of a region may contain
information about physical conditions in its past, since the chemistry often
takes significant time to respond to changes in temperature etc.
Clearly, observers of star formation need at least a basic knowledge of astrochemistry.

This paper reviews the chemistry of star-forming regions, with an emphasis on
regions of high-mass star formation. 
Section~\ref{s:basics} gives an overview of the ingredients of star-forming
matter and their interactions, both in the gas phase and in the solid phase.
Section~\ref{s:models} discusses how these processes are incorporated into
models of various physical components of star-forming regions: envelopes, hot
cores, and shocks.
Section~\ref{s:measure} reviews methods to test such models: by deriving
molecular abundances from spectroscopic observations at infrared and \smm\ 
wavelengths.
Section~\ref{s:devel} discusses recent developments in the astrochemistry of
star-forming regions, grouped into four themes: hot molecular cores, cosmic-ray
ionization, molecular depletion and deuteration, and oxygen chemistry.

Due to space limitations, this review is not complete, but biased toward the
interests of its author. 
More extensive reviews of the chemical evolution of star-forming matter are
given by \citet{vdishoeck:araa} and \citet{langer:ppiv}. For an overview of how
to extract physical parameters such as temperatures and densities from molecular
line observations, see \citet{evans:araa}. A recent review of the astrochemistry
of low-mass star formation is given by \citet{caselli:review}.

\section{Basics of astrochemistry}
\label{s:basics}

\subsection{Ingredients of star-forming matter}
\label{ss:ingredients}

%

%
The formation of stars occurs in dense interstellar clouds
with temperatures $\ltsim$100~K and densities $\gtsim$10$^3$~\ccm.
Under these conditions, the bulk species of the gas is \hh, followed by He at an
abundance of 20\% \footnote{In this paper, abundances are relative to \hh\ 
  unless noted otherwise.}.
The next most abundant species is CO, which at an abundance of
$\approx$2$\times$10$^{-4}$ \citep{vdishoeck:co} locks up all gas-phase
carbon\footnote{This paper uses words (`hydrogen') to denote elements, and
  symbols (`H') to denote their atomic form.}.
The remaining oxygen may be in the form of \hho, O or \oo, but
this issue is still open (see \S~\ref{ss:oxy}).
The main nitrogen carrier in dense clouds is also uncertain, due to the
difficulty in observing \nn\ (which has no permanent dipole moment) and N (which
has no fine structure lines).
More than 100 other molecules are known
\footnote{See \texttt{http://cdms.de} for an up-to-date list of
  molecules detected in space.},
mainly through emission lines at \smm\ wavelengths.

Besides gas, star-forming matter contains solid particles or
dust grains, at an abundance of 1\% by mass, or about 3$\times$10$^{-12}$ by
number for a typical grain radius of 0.1 \mic.
The grains consist of `cores' of amorphous carbon and silicate, covered with
layers of volatile material (`ice'), which are observed through broad absorption
features in the mid-infrared spectra of dense clouds and protostars. The
dominant ice component is \hho\ at an abundance of $\sim$10$^{-4}$; other known
ices have abundances of 1--20\% of that of \hho\ (see \citealt{vdtak:sydney} for
a list). For all ices except CO, the solid-state abundance is much higher than
can be produced in the gas phase. The data thus provide strong evidence for
grain surface reactions.


\subsection{Gas phase processes}

The rate of a gas-phase reaction between species A and B is proportional to the
number densities of A and B and to the rate coefficient $k$ of the reaction,
which can be measured in the laboratory or calculated quantum chemically.
Reactions which release energy are said to be exothermic; those which require
energy to proceed are called endothermic.
Reactions may have activation barriers even if they are net exothermic, because
for a chemical reaction to proceed, the old chemical bond must be broken before
a new one can be forged.
In addition to reactions between molecules, ions and atoms, chemical reactions
in interstellar clouds may involve electrons, dust grains, cosmic rays, and
ultraviolet photons.
The focus here is on dense clouds, defined theoretically as $n>$10$^4$~\ccm\ or
observationally as $A_V$$\gtsim$3. In such clouds, photoprocesses can be ignored
and the main molecular processes in the gas phase are:

\begin{itemize}
\item \textit{Ion-molecule reactions} such as CO $+$ \hhhp\ $\to$ HCO$^+$ $+$
  \hh, which usually do not have activation barriers and which typically occur
  at the `Langevin' rate of $\sim$10$^{-9}$ \ccms. These reactions dominate the
  chemistry of cold dark clouds. 
  Laboratory measurements indicate that some of these reactions occur at sub-Langevin
  rates, for instance the \hhhp\ $+$ HD reaction \citep{gerlich:h2d+}. However,
  scaling of the results from lab to space is not always straightforward and
  usually involves some modeling.
  
\item \textit{Dissociative recombination reactions} such as HCO$^+$ $+$ $e^-$
  $\to$ CO $+$ \hh, which have large rate coefficients of $\sim$10$^{-6}$ \ccms\ 
  due to Coulomb attraction. However, due to the low electron abundance in dense
  clouds ($\sim$10$^{-8}$), recombination cannot compete with reactions with
  \hh\ or CO if these are exothermic. Dissociative recombinations usually have
  several possible reaction products, and the `branching ratios' between these
  are often uncertain. An example are recent laboratory measurements of the
  dissociative recombination of \nnhp\ \citep{geppert:n2h+}.
  
\item \textit{Neutral-neutral reactions} usually have rate coefficients between
  10$^{-12}$ and 10$^{-10}$ \ccms, and usually (but not always) possess
  substantial activation barriers ($\sim$0.1~eV). These reactions are mainly
  important for warm, dense regions. An important example are the reactions of O
  and OH with \hh, which drive all gas-phase oxygen into \hho\ at high
  ($\gtsim$300~K) temperatures \citep{graff:oxygen}.

\end{itemize}

\subsection{Solid phase processes}

Upon collision with a grain surface, atoms and molecules may be adsorbed to form
ice mantles.
In cold, dense pre-stellar cores ($n\sim10^4$~\ccm, $T\sim10$~K), the evaporation
time for atomic species exceeds the time to scan the entire grain surface by
thermal hopping or quantum tunneling.
%
The grain surface thus acts as a catalyst for neutral-neutral reactions at low
temperatures. The reaction products are primarily \textit{saturated} species
(with mainly single chemical bonds), in marked contrast with gas-phase
ion-molecule chemistry which tends to form species with double or multiple
bonds.
The composition of the ice mantle depends primarily on the \textit{atomic}
composition of the gas phase, with a weaker dependence on grain temperature,
through differences in the volatility of atomic species \citep{tielens:hagen}.
Recent simulations \citep{chang:h2} and lab experiments \citep{liv:h2} show that
the nature of the surface also plays an important role.
The main reaction types are:
\begin{itemize}
\item At densities below 10$^4$~\ccm, H is more abundant than O, and \textit{hydrogenation}
dominates, transforming O into \hho, N into \ammo, C into CH$_4$, S into \hhs,
and CO into \hhco\ and \meth. The observed composition of interstellar ice
mantles demonstrates the dominance of this type of reaction \citep{gibb:w33a}.

\item At higher densities, O is more abundant than H and \textit{oxygenation} is
expected to become important. Support for this expectation comes from infrared
observations of ubiquitous solid \coo\ \citep{gerakines:co2}.

\item Since the atomic D/H ratio is orders of magnitude higher than the elemental
ratio, surface chemistry may also lead to \textit{deuteration}. Observational
evidence for this process is provided by the high deuterium enrichment of
evaporated \hhco\ and \meth\ in the Orion Compact Ridge \citep{charnley:meth}.
However, deuterated ice species are yet to be observed (see \S~\ref{ss:dd}). 

\end{itemize}
%


\section{Chemical models}
\label{s:models}


%

\begin{figure}
 \includegraphics[width=13cm,angle=0]{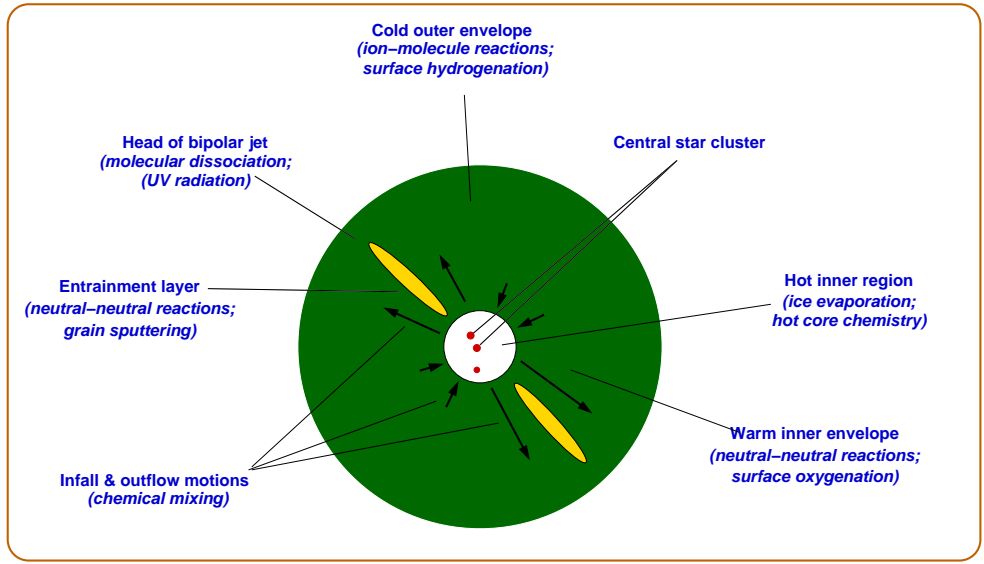}
  \caption{Schematic view of a star-forming region, with the main `chemical
    zones' indicated.}\label{f:yso}
\end{figure}

The chemistry of star-forming regions is a combination of three regimes:
`dark cloud' chemistry at $T$$\sim$10~K, `hot core' chemistry at $T$$\sim$100~K
and `shock' chemistry at $T$$\sim$1000~K. Figure~\ref{f:yso} shows the location of
these chemical zones. 

\subsection{Envelope chemistry}
\label{ss:env}

Traditional models of dark cloud chemistry follow the time evolution of an
initially atomic cloud until steady state is reached after $\sim$10$^7$~yr.
Such models typically include hundreds of species and thousands of chemical reactions.
If only gas-phase reactions are included, the observed abundances of some key
molecules such as \hho\ are not well reproduced, indicating a need to include
grain surface reactions \citep{bergin:swas}.
Since simultaneous self-consistent treatment of gas-phase and grain-surface
processes presents an enormous computational challenge, approximate methods
must be used.
%
See \citet{caselli:review} for a discussion of these approximations, and also of
the various possible starting conditions.

While the dark cloud models apply to regions of low temperature (10 -- 20~K),
evaporation of ice mantles becomes important at higher temperatures ($\gtsim$90~K).
Models of `hot core' chemistry follow the time evolution of warm gas after the
injection of large amounts of saturated molecules which have been made on dust
grain surfaces. Especially evaporated \meth\ and \hhco\ rapidly
become protonated and react with each other to form short-lived complex organic
molecules. See \citet{vdtak:sydney} for a recent review of this process.

In recent years, dark cloud models have been combined with hot core models to
describe protostellar envelopes, which have strong gradients in temperature and
density. These combined models treat the chemistry as a function of both time
and position.  The time dependence of the models may then be used to estimate
the chemical age of the region, and compare with other age estimates, for
example from dynamics.

Doty and co-workers (2002, 2004)\nocite{doty:2591,doty:16293} modeled the chemical
evolution of protostellar envelopes with centrally peaked temperature and
density distributions. They estimate the ages of these objects by calculating
molecular column densities for various times and comparing these with values
observed both in \smm\ emission and in infrared absorption.
\citet{rodgers:cocoon} and \citet{aikawa:b-e} take this approach one step
further and consider the effect of the kinematics of such envelopes on their
chemical evolution.
Very precise observations of many molecular species are needed to test these
models, which have many free parameters. For most species, agreement to factors
of 2--3 can be obtained for reasonable ages, but some species deviate much more,
indicating problems with either the initial conditions of the models or with the
assumed reaction rates.

\subsection{Shock chemistry}
\label{ss:shock}

The third type of chemistry relevant for star-forming regions is that of
interstellar shocks, which occur when powerful jets from young stars
collide with ambient material.
%
%
The chemistry of the shocked gas depends on the shock velocity.  Shocks with
velocities $\ltsim$40~\kms\ heat the ambient gas to a few 1000~K and result in
copious \hho, SO and \soo\ production in neutral-neutral reactions \citep{kaufman:c-shocks}.
In contrast, faster shocks will dissociate and ionize the gas, heat it to
10$^5$~K and produce copious ultraviolet radiation; molecules then re-form
slowly in the warm postshock gas and a largely molecule-free zone is created
\citep{hollenbach:j-shocks}. 
Such fast shocks probably occur where the ambient gas takes a direct hit from a
protostellar jet, while slower shocks take place along the sides (Fig.~\ref{f:yso}).

The grains stay cool in either type of shock, but their charge causes them to
interact with the gas. In particular, slow shocks in magnetized regions cause
ion-neutral drift, and thus to collisions of grains with \hh\ molecules. These
collisions erode the grains, liberating volatile material from their mantles and
refractory material from their cores. The enhanced abundances of \meth\ and SiO
observed towards protostellar outflows are thought to originate in such
`sputtering' \citep{bachiller:l1157}.
The recent first detections of the FeO and SiN molecules (\citealt{walmsley:feo};
\citealt{schilke:sin}) form additional tests of this scenario.

Recently, models have been developed which combine hot core chemistry with shock
chemistry, although observational tests are still inconclusive \citep{hatchell:shocks}.
Models which incorporate all three regimes (dark clouds, hot cores, and shocks)
do not exist yet, but would be valuable to describe star-forming regions.
One approach would be to develop axisymmetric models, where envelope chemistry
proceeds along the equator, and shock chemistry close to the axis.

\section{Measuring molecular abundances}
\label{s:measure}

\subsection{Observational techniques}
\label{ss:obs-tech}

The observational part of astrochemistry consists of spectroscopy of molecular
rotational lines at \smm\ wavelengths and of ro-vibrational lines at near- and
mid-infrared wavelengths. 
Either type of line may appear in emission or in absorption, depending on geometry.
However, since molecular vibrational states lie at much higher energies ($\sim$1000~K)
than rotational states ($\sim$10 -- 100~K), the excitation of \smm\ rotational lines
occurs over much larger regions than that of mid-infrared ro-vibrational lines.
In addition, absorption measurements require a background source, of which many
more are available in the mid-infrared than in the \smm.
Therefore, mid-infrared lines usually appear in absorption and \smm\ lines in emission. 
%

The advantage of absorption spectroscopy is that the angular resolution is not
limited by the size of the telescope, but by the (usually very small) size of
the background continuum source.  
On the other hand, if the lines appear in emission, they can be mapped, either
by scanning the telescope or by using array receivers.

In the mid-infrared, the molecular excitation can be measured accurately since many
ro-vibrational lines occur close together in wavelength.
Furthermore, molecules without permanent dipole moments (C$_2$H$_2$, CH$_4$,
...) can only be probed in the mid-infrared, as is the case for solid-state features.
%
%
%
On the other hand, the advantage of \smm\ spectra is that even the narrowest
spectral features can be velocity-resolved.
Furthermore, rotational lines have a much higher sensitivity
(down to abundances of $\sim$10$^{-12}$) than ro-vibrational lines.
Clearly, the infrared and \smm\ techniques are complementary, and both
are needed for a complete picture of the chemistry in star-forming regions
(e.g., \citealt{boonman:models}).

\subsection{Radiative transfer techniques}
\label{ss:radtrans}

Traditionally, molecular abundances are estimated as the ratio of the molecular
column density $N_X$ and the \hh\ column density $N_H$.
The conversion from line strength to column density requires estimates of
the molecular excitation and the line optical depth.
If several lines of the molecule have been observed, the excitation may be
estimated from the relative strengths of these lines.  If only one line has been
observed, the excitation must be estimated from observations of other species,
or from theoretical considerations.
The optical depth of an absorption line follows from the line to continuum ratio
if the profile is resolved; otherwise, the `curve of growth' method may be
applied, assuming an intrinsic line width.
For emission lines, the optical depth is usually estimated from the line strength,
assuming that the emission fills the telescope beam.
These estimates may be done analytically, or numerically using the Large
Velocity Gradient (LVG) or similar `local' approximations.
%
In the same way, $N_H$ is obtained from observations of
CO isotopic lines or dust continuum, and
the `average abundance' $N_X$/$N_H$ may
be compared with chemical models. 

The traditional procedure is appropriate for large-scale molecular clouds which
are homogeneous on the scales probed by single-dish telescopes, although the
determination of the appropriate $N_H$ often introduces considerable
uncertainty.
In star-forming regions with strong temperature and density gradients and
chemical differentiation, the procedure may be used as a first guess if little
data are available.  However, deviations from the assumptions of uniform
excitation and uniform abundance along the line of sight are considerable, and
more sophisticated methods are called for.
In recent years, the use of Monte Carlo (MC) or Accelerated Lambda Iteration
(ALI) programs has become common. Such programs take temperature, density and
velocity gradients within the source into account, and also open the possibility
to test scenarios where molecular abundances vary with location.
If enough observations of a molecule are available, an abundance \textit{profile} 
along the line of sight can be derived, rather than a mere average.
An outline of the procedure to extract such abundance profiles from multi-line
observations is given by \citet{doty:16293}.
The first results from such models include the finding of increased abundances
(`jumps') of \meth, \hhco, SO and \soo\ in the warm ($>$100~K) gas close to
young stars (Van der Tak et al.\ 2000, 2003\nocite{vdtak:meth,vdtak:sulphur};
\citealt{schoeier:16293}) which are likely due to the evaporation of icy grain
mantles. More recently, abundance decreases (`drops') of CO and \hcop\ have been
found in the cold outer parts of low-mass protostellar envelopes due to
freeze-out of CO onto the grains \citep{joergensen:inventory}.
Direct observational evidence for such differentiation usually requires high
angular resolution ($\ltsim$1$''$) as provided by interferometers. 
%
The images of the `double hot core' in the warm gas close to
one young low-mass star (\citealt{bottinelli:16293}; \citealt{kuan:16293}) and
of chemical changes along the outflow of another \citep{joergensen:iras2a}
provide two recent examples.

Both LVG and MC/ALI programs need spectroscopic and collisional input data for
the molecule that has been observed.
Until recently, such data were scattered all over the physical, chemical and
astronomical literature in many different formats.
Data for the most commonly observed molecules have now been collected into a
database \citep{schoeier:moldata}.
The associated
website\footnote{\texttt{http://www.strw.leidenuniv.nl/$\sim$moldata/}} contains
data files for many molecular and atomic species, and also a simple LVG-type
line strength calculator.


%
%

\section{Recent developments}
\label{s:devel}

\subsection{Hot core chemistry}
\label{ss:hmc}

Hot molecular cores are regions close to young stars, where ice mantles
evaporate off dust grains and high-temperature gas-phase chemistry creates
short-lived complex organic species. 
The result are very rich \smm\ spectra, for instance as observed at 800~GHz
toward Orion-KL \citep{comito:survey}. 
This work is a nice illustration of modern observing capabilities, but also of
the difficulties in assigning molecular lines in very crowded spectra.
Indeed, the claimed detection of interstellar glycine remains disputed
\citep{snyder:glycine} because of ambiguities in the line identification.

The short time scale of the hot core phenomenon ($\sim$10$^4$~yr) makes it a
possible `chemical clock', but this application requires understanding which
species are formed on the grain surface and/or in the gas phase.  Recent years
have seen progress toward this goal from several directions, both observational
and theoretical.
Ground-based high-resolution mid-infrared spectroscopy has provided the first
detection of solid $^{13}$CO towards one high-mass protostar \citep{boogert:13co}.
The differences in the absorption profiles between $^{12}$CO and $^{13}$CO
contain information about the structure of the ice matrix and its evolution
(see also Palumbo, these proceedings).
Along the same line of sight, variations of the CH$_4$ abundance were found
\citep{boogert:ch4}, which suggest that CH$_4$ is made on grains but quickly
destroyed after evaporation into the gas phase. 
Interestingly, the spectrum also shows possible absorption by solid C$_2$H$_6$
and C$_2$H$_5$OH!
The first inventories of solid state features toward low-mass protostars have
been made with the \textit{Spitzer} space observatory \citep{boogert:spitzer}.
The large \coo/\meth\ ratios suggests that the ices were formed under dense
conditions, while the band profiles suggest that thermal processing of the ice
is less important than for high-mass stars.
Mapping of solid-state features in the Serpens star-forming region
\citep{pontoppidan:ice_map} suggests that the ice formation efficiency is
strongly density dependent.


The hot core phenomenon also remains the subject of theoretical investigation,
especially the formation of complex species in the gas phase. Quantitative
understanding of these processes is still lacking, as the recent example of
methyl formate shows \citep{horn:methform}. Another recent result is that the
isomerization of \hhco\ and \meth, in particular the shifting of a D atom from
one side of the molecule to another, appears inefficient \citep{osamura:ch3od}.

\subsection{Cosmic-ray ionization}
\label{ss:zeta}

The ionization fraction of molecular clouds determines the importance of
magnetic fields for their dynamics, and sets the time scale of ion-molecule
chemistry.
In star-forming regions, which are shielded from ultraviolet radiation, 
the main source of ionization is by cosmic rays.
The bulk of the ionizations convert \hh\ into \hhhp\ (via \hh$^+$), but a small
fraction ($\approx$3\%) of cosmic rays produce He$^+$. The He ions are important
because they are the only ones capable of breaking strong chemical bonds such as
in CO or \nn.
Observations of \hhhp\ absorption and H$^{13}$CO$^+$ emission lines toward seven
young high-mass stars indicate a cosmic-ray ionization rate of
\zet$\approx$3$\times$10$^{-17}$~s$^{-1}$ \citep{vdtak:zeta}.
Very close to high-mass stars, photo-ionization plays an additional role, as the
detection of CO$^+$ and SO$^+$ toward AFGL 2591 testifies \citep{staeuber:paris}.

Recent observations indicate significant spatial variations of \zet.
Spectroscopy of \hhhp\ at 3.5 \mic\ toward a local diffuse cloud
\citep{mccall:zeta} indicates a local enhancement of \zet\ relative to the above
value by about a factor of 10, taking the distribution of species along the line
of sight into account \citep{lepetit:h3+}.
On the other hand, observations of emission lines of DCO$^+$ and other ions
toward highly shielded pre-stellar cores indicate a decrease of \zet\ by a
factor of 10 \citep{caselli:zeta}, although depletion may influence this result.
The apparent scaling of \zet\ with density may be related to magnetic fields.
Denser regions have stronger magnetic fields (Crutcher, these proceedings) and
are more able to deflect impinging cosmic rays.
Models by \citet{padoan:zeta} predict a scaling of \zet\ with density as a
result of cosmic-ray confinement by self-generated MHD waves.
More observations are needed to test these and other models.

\subsection{Depletion and Deuteration}
\label{ss:dd}

%
Another area where astrochemistry is useful are the extreme physical conditions
implied by observations of very high molecular D/H ratios in pre-stellar
cores and young protostellar envelopes.
Observations have confirmed the theoretical expectation that the strong
deuteration is linked to the depletion of CO from the gas phase by freeze-out on
dust grains \citep{bacmann:depletion}.
Current chemical models use depletion by factors of up to $\sim$10 to explain
observations of single and double deuteration (e.g.,
\citealt{roberts:depletion}), but
recent observations of triply deuterated molecules (\citealt{lis:nd3}; \citealt{vdtak:nd3};
\citealt{parise:cd3oh}; \citealt{roueff:nd3}) 
and of highly abundant \hhdp\ and \ddhp\ (\citealt{caselli:h2d+}; \citealt{vastel:d2hp})
have shown that under extreme conditions, essentially all ($>$99\%) of molecules
containing nuclei heavier than He are frozen out.
Such conditions occur at the centers of very dense pre-stellar cores on the
verge of forming stars. Very soon after stars begin to form, the heavy elements
re-appear in the gas phase and the level of deuteration goes down.
The near-absence of heavy elements from the gas phase has far-reaching consequences
for the chemistry, as shown in a series of papers by the `Deuteronomy' team
(\citealt{walmsley:depletion}; \citealt{flower:depletion}).
%
The results indicate that the abundances of \hhdp\ and \ddhp\ may be used to
constrain the grain size distribution.

The role of surface chemistry in enhancing molecular D/H ratios is uncertain,
since to date no D-bearing species are observed in the solid state 
(\citealt{dartois:hdo}; \citealt{parise:hdo}). Observed D/H enhancements around
young high-mass stars are $\sim$100$\times$ lower than around low-mass stars,
but still much higher than expected from the observed gas temperature
\citep{gensheimer:water}. In this case the D/H ratio is a `fossil' from an
earlier, colder phase, such as the recently discovered infrared dark cloud
cores. 

\subsection{Oxygen chemistry}
\label{ss:oxy}

As mentioned in \S~\ref{ss:ingredients}, the major oxygen carrier in dense
interstellar clouds is not well known. Two recent space missions (SWAS and ODIN)
have measured \smm\ lines of \hho\ and \oo\ to answer this question.
The derived \hho\ abundances range from $\sim$10$^{-4}$ in warm
($\gtsim$300~K) gas to $\sim$10$^{-8}$ in cold ($\ltsim$30~K) clouds
\citep{snell:swas}. The \oo\ molecule remains elusive, with the current best
limit on its abundance of $<$10$^{-7}$ \citep{pagani:o2}.
These observations imply that in dense interstellar clouds, \oo\ is never a
major oxygen carrier nor a major gas coolant, and \hho\ only in localized warm
regions.
The large variations in \hho\ abundance and the absence of \oo\ can be explained
by freeze-out of oxygen onto dust grains ($\ltsim$100~K), ice evaporation
($\gtsim$100~K), and neutral-neutral reactions in the gas phase ($\gtsim$250~K).

Several questions remain unanswered in this area.
The \hho\ abundances are based on a single transition, and the excitation had to
be estimated from external data. The abundances also only refer to ortho-\hho,
while most \hho\ may be in para form in cold clouds. Third, \hho\ remains
undetected in several classes of objects, notably low-mass star-forming cores.
Finally, the link between the high \hho\ abundances on small scales and the low
values on large scales is still poorly understood (e.g., \citealt{boonman:models}).
The HIFI spectrometer and PACS camera onboard the \textit{Herschel} space
observatory will allow key progress on these questions. These instruments will
carry out multi-line observations of \hho\ in $\approx$10--20$''$ beams.
For further discussion see \cite{walmsley:paris}.




\section{Conclusions and future opportunities}
\label{s:opps}

As stated in \S~\ref{s:intro}, the two uses of astrochemistry for star formation
studies are: (1) understanding which molecule traces which physical component;
(2) estimate the chemical ages of star-forming regions. The impression of this
reviewer is that the current understanding of astrochemistry is good enough for
the first goal, but not for the second. However, this situation may well change
in the next few years, when many new facilities come on-line.

New mid-infrared observations, both from the ground and with \textit{Spitzer} and
ultimately JWST, will be useful to make deeper inventories of the composition of
icy grain mantles than currently possible. In particular, the main nitrogen and
sulphur carriers are waiting to be identified, as are deuterated constituents.
Many species known from \smm\ observations are still awaiting detection by
gas-phase spectroscopy in the mid-infrared. The combination of these techniques
will be very powerful in constraining the location of these molecules (for example
\meth) and thus in building up a 3D picture of the chemical structure of
star-forming regions.

The launch of the \textit{Herschel} space observatory in 2007 will allow
spectral line surveys at Terahertz frequencies with the HIFI instrument.
These observations will considerably extend our knowledge of the gas-phase
composition of star-forming regions, particularly for light species such as
hydrides. As discussed in \S~\ref{ss:oxy}, the origin and evolution of \hho\ in
star-forming regions is another major goal of \textit{Herschel}.

Last but not least, the advent of high angular resolution at \smm\ wavelengths
with ALMA will allow us to study chemical structure on small scales. For
example, this telescope will be able to resolve regions with enhanced abundances
of, e.g., \meth\ and \soo\ in the inner envelopes of young high-mass stars.
Also of great interest is the chemistry of infrared dark cloud cores which
likely represent the initial conditions of high-mass star formation.
Clearly, the next 5--10 years will be an exciting time for the students of
astrochemistry. 

\begin{acknowledgments}
  The author thanks Malcolm Walmsley and Ted Bergin for useful discussions,
  and the organizers of the symposium for their invitation to give this review.
\end{acknowledgments}

\bibliographystyle{aa}
\bibliography{catania}




\end{document}

%% file: defs.tex
\def\hh{H$_2$}

\def\hhs{H$_2$S}
\def\hho{H$_2$O}

\def\hcop{HCO$^+$}

\def\hhco{H$_2$CO}

\def\meth{CH$_3$OH}

\def\ammo{NH$_3$}

\def\soo{SO$_2$}
\def\coo{CO$_2$}
\def\oo{O$_2$}

\def\hhhp{H$_3^+$}
\def\hhdp{H$_2$D$^+$}
\def\ddhp{D$_2$H$^+$}
\def\nnhp{N$_2$H$^+$}
\def\nn{N$_2$}

\def\zet{$\zeta_{\rm CR}$}

\def\smm{(sub-)mil\-li\-me\-ter}

\def\gtsim{{_>\atop{^\sim}}}
\def\ltsim{{_<\atop{^\sim}}}

\def\kms{km~s$^{-1}$}

\def\ccm{cm$^{-3}$}

\def\ccms{cm$^3$~s$^{-1}$}
\def\mic{$\mu$m}

\renewcommand{\citep}[1]{(\citeauthor{#1} \citeyear{#1})}